\documentclass[aps,prl,twocolumn,showpacs,superscriptaddress,groupedaddress,acknowledgment]{revtex4-1}  
\usepackage{dcolumn}   
\usepackage{bm}        

\usepackage{amsmath,amssymb,braket,fancybox}
\usepackage{amsfonts}
\usepackage{graphicx}
\usepackage{comment}
\usepackage{ascmac}
\usepackage{ascmac}
\usepackage{amsthm}
\usepackage{color}

\theoremstyle{definition}

\newtheorem*{theorem*}{Theorem}

\newtheorem*{definition*}{Definition}

\newtheorem*{proposition*}{Proposition}

\newcommand{\mc}[1]{\mathcal{#1}}
\newcommand{\mr}[1]{\mathrm{#1}}
\newcommand{\mbb}[1]{\mathbb{#1}}

\newcommand{\lrs}[1]{\left( #1 \right)}
\newcommand{\lrm}[1]{\left\{ #1 \right\}}
\newcommand{\lrl}[1]{\left[ #1 \right]}
\newcommand{\lrv}[1]{\left| #1 \right|}

\newcommand{\aln}[1]{
\begin{align}
#1
\end{align}
}

\newcommand{\ra}{\rightarrow}

\newcommand{\Tr}{\mr{Tr}}
\newcommand{\hmo}{\hat{\mc{O}}}
\begin{document}
\title{
Atypicality of Most Few-Body Observables}
\date{\today}
\author{Ryusuke Hamazaki}
\affiliation{
Department of Physics, University of Tokyo, 7-3-1 Hongo, Bunkyo-ku, Tokyo 113-0033, Japan
}
\author{Masahito Ueda}
\affiliation{
Department of Physics, University of Tokyo, 7-3-1 Hongo, Bunkyo-ku, Tokyo 113-0033, Japan
}
\affiliation{
RIKEN Center for Emergent Matter Science (CEMS), Wako 351-0198, Japan
}

\begin{abstract}
{The eigenstate thermalization hypothesis (ETH), which dictates that all diagonal matrix elements within a small energy shell be almost equal, is a major candidate to explain thermalization in isolated quantum systems.
According to the typicality argument, the maximum variations of such matrix elements should decrease exponentially with increasing the size of the system, which implies the ETH. We show, however, that the typicality argument does not apply to most few-body observables for few-body Hamiltonians when the width of the energy shell decreases at most polynomially with increasing the size of the system.}

\end{abstract}

\maketitle
\paragraph{Introduction.}
Thermalization in isolated quantum systems has long hovered over researchers~\cite{Haar55,Peres84E,Jensen85,Deutsch91,Srednicki94,Tasaki98} since von Neumann's seminal work~\cite{Neumann29}.
Recently, this problem has attracted growing interest~\cite{Polkovnikov11,Eisert15,DAlessio16} due to experimental advances in ultracold atoms~\cite{Kinoshita06,Gring12,Cheneau12,Schneider12,Trotzky12,Langen13,Schreiber15,Kaufman16,Choi16}, ions~\cite{Richerme14,Jurcevic14,Clos16,Smith16}, and superconducting qubits~\cite{Neill16}.
These experiments have motivated theorists to identify the conditions under which thermalization occurs~\cite{Reimann08,Rigol08,Santos10a,Linden09,Gogolin11,Gogolin12,Reimann15,Goldstein15,Palma15,Gogolin16,Goldstein17,Farrelly17,Anza17}.

The long-time dynamics of isolated quantum systems can be analyzed through matrix elements of an observable in the energy eigenbasis.
The eigenstate thermalization hypothesis (ETH)~\cite{Deutsch91,Srednicki94} dictates that all diagonal matrix elements within a small energy shell be almost equal~\footnote{To be precise, the ETH in this Letter is defined to hold true if $\max_{|E_\alpha-E|, |E_\beta-E|\leq \Delta E}|\mc{O}_{\alpha\alpha}-\mc{O}_{\beta\beta}|/||\mc{\hat{O}}||_\mr{op}$ vanishes in the thermodynamic limit for a subextensive $\Delta E$. Since this definition does not require any detailed structure of matrix elements within the energy shell, it is usually weaker than another definition used in Ref.~\cite{Beugeling14} or the ansatz in Ref.~\cite{Srednicki99}. Nevertheless, our definition has a clear physical meaning that it gives a sufficient condition for any initial system with a subextensive standard deviation of energy to thermalize. Consequently, many previous papers use the definition similar to ours~\cite{Neumann29,Goldstein10L,Reimann15,Palma15}.}. Then the expectation value of an observable in the steady state can be calculated from the microcanonical ensemble for all initial states with small energy fluctuations~\footnote{Note that a seminal idea of  thermalization due to the ETH was already present in Ref.~\cite{Neumann29} in what von Neumann called macrospaces.}.
Meanwhile, off-diagonal matrix elements characterize autocorrelation functions and temporal fluctuations~\cite{DAlessio16}.
It is thus of fundamental importance to understand how such matrix elements generically behave in  macroscopic systems.

To be specific, consider a set of eigenstates $\{\ket{E_\alpha}\}$ of the Hamiltonian and introduce a projector $\hat{\mc{P}}_\mr{sh}=\sum_{|E_\alpha-E|\leq \Delta E}\ket{E_\alpha}\bra{E_\alpha}$ onto the Hilbert space $\mc{H}_\mr{sh}$ for an energy shell of median $E$ and width $2\Delta E$. 
Let the spectral decomposition of an observable $\hmo$ projected onto $\mc{H}_\mr{sh}$ be $\hat{\mc{P}}_\mr{sh}\hmo\hat{\mc{P}}_\mr{sh}=\sum_{i=1}^{d_\mr{sh}}a_i\ket{a_i}\bra{a_i}$, where $d_\mr{sh}=\dim[\mc{H}_\mr{sh}]$ is the dimension of the Hilbert space within the energy shell. 
Then the matrix elements of $\mathcal{\hat{O}}$ within the energy shell can be expressed as $\mc{O}_{\alpha\beta}=\braket{E_\alpha|\hat{\mc{P}}_\mr{sh}\hmo\hat{\mc{P}}_\mr{sh}|E_\beta}=\sum_{i}a_iU_{\alpha i}U^*_{\beta i}$, where $U_{\alpha i}:=\braket{E_\alpha|a_i}$ constitutes the $d_\mr{sh}\times d_\mr{sh}$ unitary matrix $U$.

{To investigate the ETH, let us consider the maximum variation of $\mc{O}_{\alpha\alpha}$ within the energy shell.
This quantity enables us to directly judge whether all diagonal matrix elements are almost equal, which is required for justifying thermalization from arbitrary initial states~\footnote{This type of the ETH is often called as the strong ETH. On the other hand, if the variance of $\mc{O}_{\alpha\alpha}$ within the shell vanishes in the thermodynamic limit, the system is said to satisfy the weak ETH. It is known that the weak ETH does not imply thermalization~\cite{Biroli10}.}. }
As shown in Ref.~\cite{Reimann15} (see Appendix I of the Supplemental Material~\footnote{See Supplemental Material for detailed derivations, numerical calculations, and generalization of the results, which includes Ref.~\cite{Goldstein10}}), the maximum deviation of $\mc{O}_{\alpha\alpha}$ from its average value decreases exponentially with increasing the size of the system for almost all (typical) $U$'s over the unitary Haar measure. 
This mathematical property is referred to as the typicality with respect to the unitary Haar measure~\footnote{This definition is different from that of the typicality of pure states~\cite{Goldstein06,Popescu06,Reimann07,Tasaki16}}.
Based on the typicality, it is argued~\cite{Reimann15} that for actual $\hat{H}$ and $\hmo$ of our concern the variations of $\mc{O}_{\alpha\alpha}$ are exponentially small.
We refer to this conjecture as the typicality argument~\cite{Reimann15} to distinguish it from the above-mentioned (mathematically rigorous) typicality.
Since exponentially small variations of $\mc{O}_{\alpha\alpha}$ imply the ETH within the same energy shell,
the typicality argument offers a possible scenario for the justification of the ETH~\footnote{Note that their argument essentially uses the same measure for matrix elements as ours to quantify the ETH (see also Appendix I).}.
Such an idea was originally put forth by von Neumann for macrospaces~\cite{Neumann29,Goldstein10L} and it has recently been generalized to arbitrary observables~\cite{Rigol12,Reimann15}.
{
Note that the spirit of the typicality argument is similar to that of applying random matrix theory (RMT)~\cite{Wigner51,Brody81,Bohigas84,Goldstein10L,Reimann15} to physics~\footnote{Some of the spectral statistics in semiclassically chaotic or disordered systems are proven to have the universality predicted by RMT using the nonlinear sigma model or the periodic-orbit theory~\cite{Mirlin00,Muller04,Haake}}. 
}


In this Letter, however, we show that the typicality argument cannot be applied to most few-body observables for lattice Hamiltonians.
In fact, we show that diagonal matrix elements for most few-body observables do not behave typically even if the energy width decreases algebraically with increasing the size of the system.
In other words, the maximum variation of $\mc{O}_{\alpha\alpha}$ does not decrease exponentially.
Our approach provides rigorous results without assuming the unitary Haar measure~\cite{Reimann15} nor the specific form of matrix elements proposed in Ref.~\cite{Srednicki99}.

\paragraph{Setup.}
We assume that the energy width $\Delta E$ scales with the system size $N$ as $\Delta E \propto N^{-p}$ and that $d_\mr{sh}$ increases exponentially with $N$,
where $-1<p<0$ (sub-extensive) for the energy width of the microcanonical ensemble and $p=\frac{2}{\mc{D}}$ for that of the diffusive energy (many-body Thouless energy)~\cite{DAlessio16} with $\mc{D}$ being the spatial dimension.

\begin{figure}
\begin{center}
\includegraphics[width=9cm]{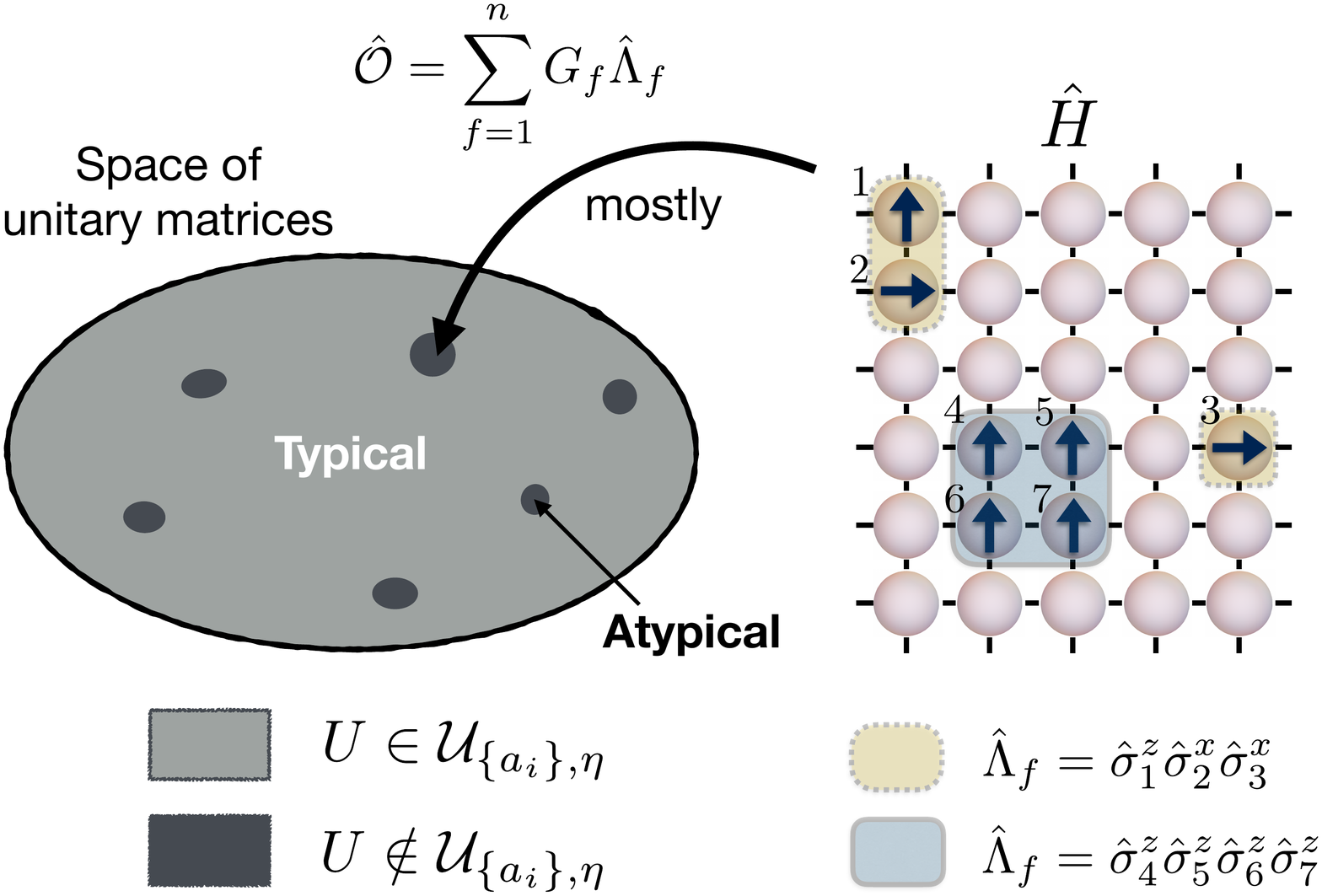}
\caption{
(Left) Space of all the unitary matrices $U$'s whose matrix elements are constituted from inner products between the eigenbases of the Hamiltonian of a system and those of an observable.
Almost all $U$'s with respect to the unitary Haar measure belong to $\mc{U}_{\{a_i\},\eta}$, {a set of all unitary matrices for which} the maximum variations of matrix elements $\mc{O}_{\alpha\alpha}$ decrease exponentially with increasing the size of the system.
(Right) In a system with few-body interactions, we consider few-body observables $\hmo$ that are expressed as random linear combinations of few-body operator bases $\{\hat{\Lambda}_f\}$.
Then, for most realizations of $\hmo$, the corresponding $U$'s are atypical when the energy window under consideration decreases at most polynomially with increasing the size of the system.
}
\label{typpp}
\end{center}
\end{figure}

In the following we consider a spin system on $N$ lattice sites.
The entire Hilbert space can be written as $\mc{H}=\bigotimes_{x=1}^N\mc{H}_x$, where $\mc{H}_x$ is a local Hilbert space at site $x$.
Let $\mc{L}(\mc{H})$ and $\mc{L}(\mc{H}_x)$ be operator spaces acting on $\mc{H}$ and $\mc{H}_x$, respectively.
We take an orthonormal basis set for $\mc{L}(\mc{H}_x)$ as $\lrm{\hat{\lambda}_x^0:=\hat{\mbb{I}}_x, \hat{\lambda}_x^{1},\cdots,\hat{\lambda}_x^{S^2-1}}$, where $S=\dim[\mc{H}_x]$ and $\hat{\lambda}^\mu_x\:(0\leq \mu\leq S^2-1)$ are $S\times S$ Hermitian matrices subject to $\Tr_x[ \hat{\lambda}_x^{\mu} \hat{\lambda}_x^{\mu'}]=S\delta_{\mu{\mu'}}$.
Then, the basis set that spans $\mc{L}(\mc{H})$ is written as $\mc{B}_N=\lrm{\hat{\Lambda}'_{\mu_1,\cdots,\mu_N}=\bigotimes_{x=1}^N\hat{\lambda}_x^{\mu_x}|{0\leq\mu_x\leq S^2-1}}$, where $\Tr[\hat{\Lambda}'_{\mu_1,\cdots,\mu_N}\hat{\Lambda}'_{\mu'_1,\cdots,\mu'_N}] =S^N\prod_{x=1}^N\delta_{\mu_x\mu_x'}$.

We next define $m$-body operators.
For this purpose, we take a basis set $\mc{B}_m\subset \mc{B}_N$ whose elements act nontrivially on at most $m$ sites: 
$\mc{B}_m=
\lrm{\bigotimes_{i=1}^q\hat{\lambda}_{x_i}^{\alpha_{x_i}}|1\leq q\leq m,1\leq x_i\leq N, 1\leq\alpha_{x_i}\leq S^2-1}$
for $m\geq 1$ and $\mc{B}_0=\lrm{\bigotimes_{x=1}^N\hat{\lambda}_{x}^{0}}$.
Then $m$-body operators are defined as a linear combination of elements in $\mc{B}_m$ but not in $\mc{B}_{m-1}$.
If $m\:(m\ll N)$ does not depend on $N$, we call them few-body operators.
We note that our few-body operators are defined in a much broader sense than usual.

To discuss characteristic behaviors of few- and many-body observables, we next consider observables which are randomly chosen from at most $m$-body operators.
\begin{definition*}[Randomly chosen observables from $\mc{L}_m$]
Let $\hat{\Lambda}_1,\cdots,\hat{\Lambda}_n$ be elements in $\mathcal{B}_m$, where $n=\sum_{q=0}^m\frac{N!}{q!(N-q)!}(S^2-1)^q$ is the number of the bases and  $\Tr[\hat{\Lambda}_{f}\hat{\Lambda}_{g}]=S^N\delta_{fg}$.
Let us consider a set $\mc{L}_m$ of at most $m$-body observables, which can be written as a linear combination of $\hat{\Lambda}_f$.
Now, we take an observable $\hat{G}\in\mc{L}_m$ expressed as
\aln{
\hat{G}=\sum_{f=1}^{n}G_f\hat{\Lambda}_f,
}
where real variables $\vec{G}=(G_1,\cdots, G_f,\cdots,G_{n})$ are randomly chosen according to an arbitrarily given probability distribution $P(\vec{G})$.
When $P(\vec{G})$ is invariant under an arbitrary $n\times n$ orthogonal transformation, we call $\hat{G}$ an observable randomly chosen from $\mc{L}_m$~\footnote{This definition depends on the choice of $\{\hat{\Lambda}_f\}$ and $P(\vec{G})$. Our discussion holds true for any choice as long as the orthonormality, Hermiticity of $\{\hat{\Lambda}_f\}$ and the invariance of $P(\vec{G})$ under orthogonal transformations hold true}.
Note that we may arbitrarily choose $P(\vec{G})$ to suit our purpose; in contrast, if we choose $U$ from a unitary Haar measure, it is unclear from what probability distribution an observable is chosen. In this sense, our scheme of sampling observables has a well-defined operational meaning.
\end{definition*}

\paragraph{Atypicality of most few-body observables.}
We investigate the behavior of matrix elements of random observables defined above and compare it with what  the Haar measure predicts.
As defined above, the diagonal matrix elements $\mc{O}_{\alpha\alpha}$ within the energy shell $(|E_\alpha-E|\leq \Delta E)$ are given by $\mc{O}_{\alpha\alpha}=\sum_{i=1}^{d_\mr{sh}}a_i|U_{\alpha i}|^2$.
{We define} $\mc{U}_{\{a_i\},\eta}$ {as} a set of {all} $U$'s that lead to the inequality $\max_{|E_\alpha-E|, |E_\beta-E|\leq \Delta E}\lrv{\mc{O}_{\alpha\alpha}-\mc{O}_{\beta\beta}}\leq ||\hmo||_\mr{op} d_\mr{sh}^{-\eta}$ for given ${\{a_i\}}$, where $||\hmo||_\mr{op}$ denotes the operator norm and $\eta>0$.
This inequality means that the maximum variation of $\frac{\mc{O}_{\alpha\alpha}}{||\hmo||_\mr{op}}$ within the energy shell decreases exponentially as a function of $N$, which also implies the ETH of $\frac{\hmo}{||\hmo||_\mr{op}}$.
As illustrated in Fig. \ref{typpp}, almost all (typical) $U$'s with respect to the Haar measure belong to $\mc{U}_{\{a_i\},\eta}$ in the thermodynamic limit for $0<\eta<\frac{1}{2}$ (see Appendix I for the proof~\footnotemark[4]).

We first consider a few-body Hamiltonian (i.e., Hamiltonian with few-body interactions) and few-body observables.
We show that for most few-body observables, the corresponding $U$ is atypical in the sense that $U\notin \mc{U}_{\{a_i\},\eta}$ (see Fig. \ref{typpp}).
In fact, we can show the following theorem.
\begin{theorem*}[Atypicality of most few-body observables]
Let us consider a $k$-body Hamiltonian, and assume that $N$ is sufficiently large and that $m \:(k\leq m\ll N)$ is independent of $N$.
Suppose that we randomly choose an observable {$\hmo=\sum_fG_f\hat{\Lambda}_f$} from $\mc{L}_m$, from which we obtain the corresponding ${\{a_i\}}$  and $U$.
Then,
\aln{\label{mthm}
\mbb{P}_{\mc{L}_m}[U\in \mc{U}_{\{a_i\},\eta}]\leq  \frac{\sqrt{\pi n}||\hat{H}||_\mr{op}\Lambda}{2\Delta E}\frac{\Gamma\lrs{\frac{n}{2}}}{\Gamma\lrs{\frac{n-1}{2}}} d_\mr{sh}^{-\eta},
}
where $\mbb{P}_{\mc{L}_m}$ denotes the probability with respect to $P(\vec{G})$, and
 $\Lambda=\max_f||\hat{\Lambda}_f||_\mr{op}\leq S^{\frac{m}{2}}$.
When $||\hat{H}||_\mr{op}$ does not grow exponentially with respect to $N$, the left-hand of the inequality (\ref{mthm}) vanishes for large $N$.
Note that the assumption of the scaling $\Delta E\propto N^{-p}$ is sufficient to bound the right-hand side.
\end{theorem*}

The inequality (\ref{mthm}) shows that, for physically relevant Hamiltonians and most few-body observables, 
$\max_{|E_\alpha-E|, |E_\beta-E|\leq \Delta E}\lrv{\mc{O}_{\alpha\alpha}-\mc{O}_{\beta\beta}}$ does not decrease as a power of ${d}_\mr{sh}$.
This means that the corresponding $U$ is atypical.
As long as $m$ satisfies $k\leq m$ and and is independent of $N$ (i.e., few-body), atypicality holds true for every $m$.

\begin{figure}
\begin{center}
\includegraphics[width=9cm]{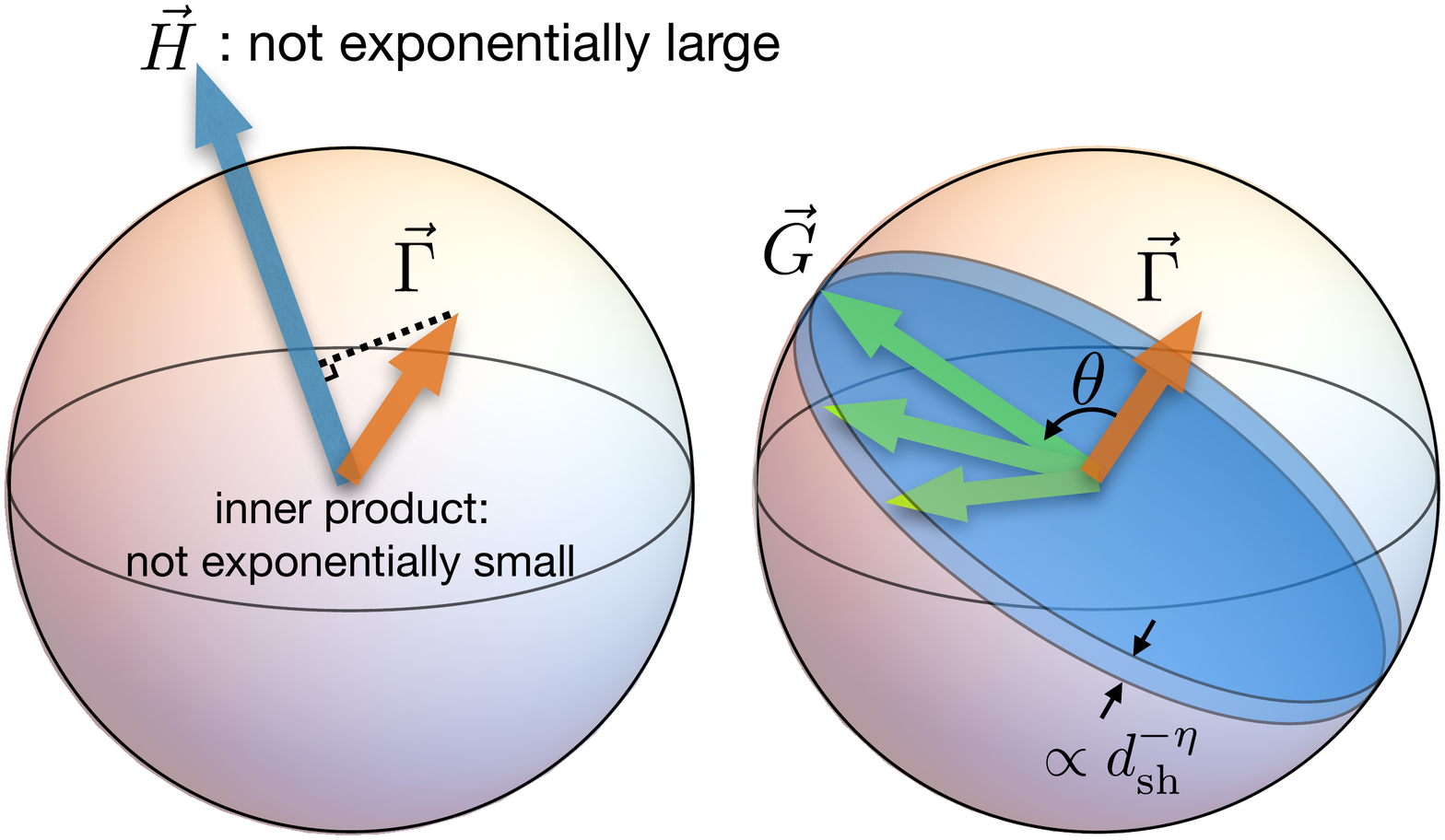}
\caption{
Schematic illustration of the idea behind the proof of \textbf{Theorem.} 
(Left) We first show that $|\vec{\Gamma}|$ does not decrease exponentially with increasing $N$ unless $|\vec{H}|$ is exponentially large.
This can be seen from the fact that $|\vec{\Gamma}\cdot\vec{H}|$ does not decrease exponentially as a function of $N$.
(Right) Then, for $|\vec{\Gamma}\cdot \vec{G}|$ to be exponentially small ($\leq ||\hat{\mc{O}}||_\mr{op} d_\mr{sh}^{-\eta}$), $\vec{G}$ should be almost orthogonal to $\vec{\Gamma}$ (we assume that $|\vec{G}|$ is not exponentially large).
The probability of such an event is exponentially small ($\propto d_\mr{sh}^{-\eta}$) unless the dimension $n$ of the hypersphere is exponentially large.
This is the case for few-body observables.
}
\label{atyppp}
\end{center}
\end{figure}

\textit{Proof of \textbf{Theorem}} (\textit{see Fig. \ref{atyppp}}).
\:We first note that $\hat{H}\in\mc{L}_m$ satisfies the following condition for a $k$-body Hamiltonian:
\aln{\label{atyp}
{\max_{|E_\alpha-E|, |E_\beta-E|\leq \Delta E}|(\hat{H})_{\alpha\alpha}-(\hat{H})_{\beta\beta}|}= \xi_\mr{d}.
}
Here $\xi_\mr{d}=2\Delta E$ does not decrease faster than polynomial in $N$.

Let $\gamma$ and $\delta$ be labels of eigenstates that satisfy $(\hat{H})_{\gamma\gamma}-(\hat{H})_{\delta\delta}= \xi_\mr{d}$.
Define $\Gamma_f=(\hat{\Lambda}_f)_{\gamma\gamma}-(\hat{\Lambda}_f)_{\delta\delta}$.
Then the expansion
$\hat{H}=\sum_{f=1}^nH_f\hat{\Lambda}_f$
leads to
$\vec{H}\cdot\vec{\Gamma}= \xi_\mr{d}$,
where $\vec{H}=(H_1,\cdots,H_{n})$ and $\vec{\Gamma}=(\Gamma_1,\cdots,\Gamma_{n})$.
Since $|\vec{H}|=\sqrt{\frac{\Tr[\hat{H}^2]}{S^N}}\leq ||\hat{H}||_\mr{op}$, we obtain
\aln{
|\vec{\Gamma}|\geq \frac{\xi_\mr{d}}{||\hat{H}||_\mr{op}}.
}

Next, we evaluate the left-hand side of Eq. (\ref{mthm}).
Since $\max_{|E_\alpha-E|, |E_\beta-E|\leq \Delta E}\lrv{\mc{O}_{\alpha\alpha}-\mc{O}_{\beta\beta}}\geq |\vec{G}\cdot\vec{\Gamma}|$, we obtain
\aln{\label{atosukoshi}
\mbb{P}_{\mc{L}_m}&\lrl{ \max_{|E_\alpha-E|, |E_\beta-E|\leq \Delta E}\lrv{\mc{O}_{\alpha\alpha}-\mc{O}_{\beta\beta}}\leq ||\hmo||_\mr{op}\epsilon}\nonumber\\
&\leq \mbb{P}_{\mc{L}_m}\lrl{|\vec{G}\cdot\vec{\Gamma}| \leq ||\hmo||_\mr{op}\epsilon}
}
for any $\epsilon>0$, where we use the fact $\mbb{P}[a\leq c] \geq\mbb{P}[b\leq c]$ for $a \leq b$.

To evaluate Eq.~(\ref{atosukoshi}), note that the probability $P(\vec{G})d\vec{G}$ can be written as $P'(|\vec{G}|)|\vec{G}|^{n-1}d|\vec{G}|d\Omega$ because of the invariance under orthogonal transformations ($\Omega$ denotes the high-dimensional solid angle).
Then, denoting the angle between $\vec{G}$ and $\vec{\Gamma}$ by $\theta$, we obtain
\aln{\label{owari}
 \mbb{P}_{\mc{L}_m}&\lrl{|\vec{G}\cdot\vec{\Gamma}| \leq ||\hmo||_\mr{op}\epsilon}\nonumber\\
&\leq \mbb{P}_{\mc{L}_m}\lrl{|\cos\theta| \leq \frac{\sqrt{n}||\hat{H}||_\mr{op}\Lambda\epsilon}{\xi_\mr{d}} }\nonumber\\
&\leq \frac{\sqrt{\pi n}||\hat{H}||_\mr{op}\Lambda\epsilon}{2\Delta E}\frac{\Gamma\lrs{\frac{n}{2}}}{\Gamma\lrs{\frac{n-1}{2}}}.
}
Here, in deriving the second line, we use
$
||\hmo||_\mr{op}\leq\Lambda |\vec{G}|\sqrt{n}
$
that results from the property of an operator norm and  the Cauchy-Schwartz inequality (see Eq. (37) in Supplementary Material~\footnotemark[4]).
{
With $\epsilon=d_\mr{sh}^{-\eta}$, the left-hand side of (\ref{atosukoshi}) becomes $\mbb{P}_{\mc{L}_m}[U\in \mc{U}_{\{a_i\},\eta}]$, which, together with (\ref{owari}), completes the proof of the theorem.}\qed

{
We note that our \textbf{Theorem} holds true for an arbitrary $k$-body Hamiltonian. 
Thus, it is natural to take $\hat{H}$ as a Hamiltonian written as a sum of spatially local operators (such locality is expected to be necessary for standard statistical mechanics).
In this case, $||\hat{H}||_\mr{op}\propto N$ is expected.
Thus, for the left-hand of (\ref{atosukoshi}) to vanish, we can take $\epsilon\propto N^{-z}\sim n^{-\frac{z}{m}}\:(z>1+p+m)$ if $\Delta E\propto N^{-p}$, since $\sqrt{n}\Lambda\frac{\Gamma\lrs{\frac{n}{2}}}{\Gamma\lrs{\frac{n-1}{2}}}\ra n\sim N^m$ for large $N$. This means that the maximum variation for most few-body observables decays slower than $\sim N^{-z}\sim n^{-\frac{z}{m}}\:(z>1+p+m)$.
}

{
We emphasize that observables treated in the theorem can be either
local or non-local observables as long as they are few-body.
Thus, our theorem applies to momentum distributions~\cite{Rigol08,Rigol12} and structure factors~\cite{TorresHerrera14}, which are non-local but nevertheless expected to obey standard statistical mechanics~\cite{DAlessio16}. 
Our theorem implies that it is likely that $U$ satisfies $U\notin\mc{U}_{\{a_i\},\eta}$ for local (or non-local but physical) observables, unless special reasons dictate otherwise.
However, it is often the case that we are specifically interested in spatially local observables.
An extension of our theorem to most local observables for translation-invariant Hamiltonians with local interactions is given in Appendix IV~\footnotemark[4].
}

\paragraph{A measure consistent with the typicality argument.}
As already noted, \textbf{Theorem} is meaningful only when $m$ does not grow in comparable proportion with $N$.
Indeed, we can show that for most observables randomly chosen from $\mc{L}_{m=N}$, the correponding $U$'s satisfy $U\in \mc{U}_{\{a_i\},\eta}$.
This means that we can construct an operational measure consistent with the typicality argument if $N$-body observables are available, as stated in the following proposition.

\begin{proposition*}[]
Consider randomly choosing an observable {$\hmo=\sum_fG_f\hat{\Lambda}_f$} from $\mc{L}_N$, from which we obtain the corresponding ${\{a_i\}}$  and $U$.
Then, we can show that (see Appendix II~\footnotemark[4])
\aln{\label{zenbu}
\mbb{P}_{\mc{L}_N}[U\notin \mc{U}_{\{a_i\},\eta}]\leq  2d\exp\lrl{-\frac{dd_\mr{sh}^{-2\eta}}{72\pi^3}},
}
where $d=\dim[\mc{H}]=S^N$.
The right-hand side vanishes for sufficiently large $N$ when $\eta<\frac{1}{2}$.
\end{proposition*}

This proposition suggests that most random observables chosen from $\mc{L}_N$ satisfy the ETH within the energy shell under the normalization $||\hmo||_\mr{op}=1$.
In Appendix III~\footnotemark[4], we show numerical results indicating that many-body correlations can satisfy the ETH (a similar numerical result was presented in Ref.~\cite{Hosur16}).

We note that we can make similar analyses for off-diagonal matrix elements $\mc{O}_{\alpha\beta}\:(E_\alpha\neq E_\beta)$, which are related to the Fourier transform of the autocorrelation function~\cite{Khatami13,Beugeling15,Luitz16}.
In Appendix V~\footnotemark[4], we show that the magnitudes of off-diagonal matrix elements for most few-body observables fluctuate within the energy shell more than what the uniform Haar measure predicts.
This can be proven from the fact that an operator written in the form of $i[\hat{A},\hat{H}]$ has atypical off-diagonal matrix elements.
On the other hand, we can construct an operational measure consistent with the typicality argument if many-body observables are available.

We briefly comment on previous investigations on whether the typicality (or RMT) argument applies to a realistic setup.
They mainly investigate the behavior of $U$ in light of the complexity of the Hamiltonian, motivated by analyses in semiclassical systems~\cite{Berry77R,Srednicki94}.
When the Hamiltonian commutes with many local conserved quantities due to, e.g.,  many-body localization~\cite{Basko06,Pal10}, matrix elements of few-body observables are not typical (i.e., atypical) ~\cite{Rigol09,Ikeda13,Steinigeweg13,Beugeling14,Beugeling15,Alba15,Hamazaki16G,Luitz16,Lan17}.
In contrast, the typicality argument has been conjectured to be applicable to a generic nonintegrable Hamiltonian~\footnote{As explicit counterexamples, some nonintegrable systems have been proposed for which the ETH does not hold even without localization~\cite{Hamazaki16G,Shiraishi17,Mori17}.} and few-body observables, for which the matrix elements are expected to be calculated by RMT within a sufficiently small energy shell~\cite{DAlessio16}.
{
For some statistics (e.g., variances of matrix elements), this conjecture has been numerically tested in Refs.~\cite{Khatami13,Steinigeweg13,Beugeling14,Beugeling15,Mondaini17}.
On the other hand, our results show that {the Haar measure cannot predict} the maximum variation of $\mc{O}_{\alpha\alpha}$ for most few-body or local observables in macroscopic systems (the latter requires a translation-invariant $\hat{H}$), when the width of the energy shell decreases at most polynomially with increasing the size of the system.}

\paragraph{Conclusions and discussions.}
We have reexamined the typicality argument that relies on the unitary Haar measure by focusing on few-body observables.
By considering an arbitrary few-body Hamiltonian {(which can be local)} and random few-body observables,
we have shown that matrix elements do not behave typically for most few-body observables even if the energy width decays algebraically with increasing the size of the system (\textbf{Theorem}).
We have also constructed an operational measure consistent with the typicality argument on diagonal matrix elements (\textbf{Proposition}).
This is possible if many-body observables are available.

Our approach provides rigorous results without assuming any specific form of matrix elements.
In fact, if we assume that all $\mc{O}_{\alpha\alpha}$'s are written as $\mc{A}(E_\alpha)$ with a smooth function $\mc{A}$ of energy in the thermodynamic limit~\cite{Srednicki99} and that $\frac{d\mc{A}(E)}{dE}$ is not exponentially small, the atypicality of diagonal matrix elements is expected.
Namely, under such assumptions the maximum deviation will be $\sim \frac{d\mc{A}(E)}{dE}\times 2\Delta E$ in the thermodynamic limit, which is not exponentially small if $\Delta E\propto N^{-p}$.
However, our proof of \textbf{Theorem} does not rely on these assumptions.
Moreover, \textbf{Theorem} and \textbf{Proposition} show that the few-body property of observables is crucial in considering statistics of matrix elements, which was not addressed in previous literature~\footnote{In other words, our work rigorously proves and refines the observation in Ref.~\cite{DAlessio16} that the prediction of the RMT seems inapplicable for a non-small energy shell.}.
Our results suggest that the above assumption for $\frac{d\mc{A}(E)}{dE}$ often seems to hold in numerics~\cite{Rigol09,Kim14} because few-body observables are mainly concerned~\footnote{We note that in Floquet nonintegrable systems, where the energy is not conserved and our theorem does not apply, $\frac{d\mc{A}(E)}{dE}$ ($E$ represents the quasienergy) is found to be small in general even for few-body observables~\cite{DAlessio14,Lazarides14}.}.

Our results indicate that the typicality argument based on the Haar measure does not apply to realistic Hamiltonians and most few-body observables, if the width of the energy shell decreases at most polynomially.
For the diagonal matrix elements, the typicality argument (see Appendix I for detail~\footnotemark[4]) cannot be used to justify the ETH.
{
However, we have not excluded the possibility that the maximum variation of diagonal matrix elements decreases algebraically with increasing the size of the system. If this is the case, the ETH still holds true and thermalization occurs in the thermodynamic limit.}
We also note that we cannot judge the validity of von Neumann's original argument~\cite{Neumann29} on the basis of the present study.
In fact, he took a coarse-grained procedure of the original macroscopic observables, which adds subextensive corrections to these observables.
Such corrections are negligible for discussing thermalization in macroscopic systems, but make the inequality (\ref{mthm}) inapplicable.

{
While we have mainly considered the maximum variation of diagonal matrix elements to investigate whether the typicality argument can explain the ETH, our technique of random few-body observables may be applied to investigate other properties of thermalization.
In particular, it is worthwhile to investigate whether the equilibration timescales of generic few-body observables are explicitly shown to differ from what the Haar measure predicts~\cite{Reimann16}.
}

\begin{acknowledgments}
We are grateful to Zongping Gong and Tomohiro Shitara for reading the manuscript with valuable comments.
We also thank Marcos Rigol and Takahiro Sagawa for fruitful discussions and comments. 
This work was supported by KAKENHI Grant No. JP26287088 from the Japan Society for the Promotion of Science, a Grant-in-Aid for Scientific Research on Innovative Areas “Topological Materials Science” (KAKENHI Grant No. JP15H05855), and the Photon Frontier Network Program from MEXT of Japan,
and the Mitsubishi Foundation.
R. H. was supported by the Japan Society for the Promotion of Science through Program for Leading Graduate Schools (ALPS) and JSPS fellowship (JSPS KAKENHI Grant No. JP17J03189).
\end{acknowledgments}

\bibliography{../../../../refer_them}

\end{document}


\title{
Supplementary Material for ``Atypicality of Most Few-Body Observables"}
\date{\today}
\author{Ryusuke Hamazaki}
\affiliation{
Department of Physics, University of Tokyo, 7-3-1 Hongo, Bunkyo-ku, Tokyo 113-0033, Japan
}
\author{Masahito Ueda}
\affiliation{
Department of Physics, University of Tokyo, 7-3-1 Hongo, Bunkyo-ku, Tokyo 113-0033, Japan
}
\affiliation{
RIKEN Center for Emergent Matter Science (CEMS), Wako 351-0198, Japan
}

\pacs{05.30.-d, 03.65.-w}

\maketitle

\section{Typicality of diagonal matrix elements}
{We begin by recalling that diagonal matrix elements $\mc{O}_{\alpha\alpha}$ within the energy shell $(|E_\alpha-E|\leq \Delta E)$ are determined by a unitary matrix $U$ through $\mc{O}_{\alpha\alpha}=\sum_{i=1}^{d_\mr{sh}}a_i|U_{\alpha i}|^2$, if eigenvalues $\{a_i\}$ are fixed.}
{We define}
 $\mc{U}_{\{a_i\},\eta}$ {as}
 a set of {all} $U$'s that lead to the inequality $\max_{|E_\alpha-E|, |E_\beta-E|\leq \Delta E}\lrv{\mc{O}_{\alpha\alpha}-\mc{O}_{\beta\beta}}\leq ||\hmo||_\mr{op} d_\mr{sh}^{-\eta}$ for given ${\{a_i\}}$. Here $||\hmo||_\mr{op}$ denotes an operator norm and $\eta>0$.
This inequality means that the maximum difference among $\frac{\mc{O}_{\alpha\alpha}}{||\hmo||_\mr{op}}$ within the energy shell is exponentially small in $N$, which also implies the ETH of $\frac{\hmo}{||\hmo||_\mr{op}}$.

There have been a number of attempts~\cite{Neumann29,Goldstein10,Goldstein10L,Reimann15}  to justify the ETH by proving $U\in\mc{U}_{\{a_i\}}$ for almost all $U$'s with respect to the Haar measure.
We present here a slightly modified version of Reimann's result~\cite{Reimann15} to compare it with our results in the main text:
\aln{\label{ineq}
\mbb{P}_U[U\notin \mc{U}_{\{a_i\},\eta}]\leq 2d_\mr{sh}\exp\lrl{-\frac{d_\mr{sh}^{1-2\eta}}{72\pi^3}},
}
 where $\mbb{P}_U$ denotes a probability distribution with respect to the unitary Haar measure.

\textit{Proof.}\:
In Ref.~\cite{Reimann15}, Reimann has shown that (see Eq. (40) in the supplementary material of the reference)
\aln{
\mbb{P}_U\lrl{\max_{|E_\alpha-E|\leq \Delta E}\lrv{\mc{O}_{\alpha\alpha}-\mc{O}_\mr{sh}}> \epsilon}\leq 2d_\mr{sh}\exp\lrl{-\frac{2\epsilon^{2}d_\mr{sh}}{9\pi^3\Delta_\mc{O}^2}},
}
where $\Delta_\mc{O}=\max_ia_i-\min_ia_i$ and $\mc{O}_\mr{sh}=\frac{1}{d_\mr{sh}}\sum_{|E_\alpha-E|\leq \Delta E}\mc{O}_{\alpha\alpha}$.
Since 
\aln{
\frac{1}{2}\max_{|E_\alpha-E|, |E_\beta-E|\leq \Delta E} \lrv{\mc{O}_{\alpha\alpha}-\mc{O}_{\beta\beta}}\leq \max_{|E_\alpha-E|\leq \Delta E}\lrv{\mc{O}_{\alpha\alpha}-\mc{O}_\mr{sh}},
}
 we obtain
\aln{\label{sikidayo}
\mbb{P}_U\lrl{\max_{|E_\alpha-E|, |E_\beta-E|\leq \Delta E}\lrv{\mc{O}_{\alpha\alpha}-\mc{O}_{\beta\beta}}> 2\epsilon}
\leq
\mbb{P}_U\lrl{\max_{|E_\alpha-E|\leq \Delta E}\lrv{\mc{O}_{\alpha\alpha}-\mc{O}_\mr{sh}}> \epsilon},
}
where we use the fact that $\mbb{P}[a> c] \leq\mbb{P}[b> c]$ for $a \leq b$.
Substituting $2\epsilon=||\hmo||_\mr{op} d_\mr{sh}^{-\eta}$ into (\ref{sikidayo}), we obtain 
\aln{
\mbb{P}_U\lrl{\max_{|E_\alpha-E|, |E_\beta-E|\leq \Delta E}\lrv{\mc{O}_{\alpha\alpha}-\mc{O}_{\beta\beta}}> ||\hmo||_\mr{op} d_\mr{sh}^{-\eta}}&\leq
\mbb{P}_U\lrl{\max_{|E_\alpha-E|\leq \Delta E}\lrv{\mc{O}_{\alpha\alpha}-\mc{O}_\mr{sh}}> \frac{1}{2}||\hmo||_\mr{op} d_\mr{sh}^{-\eta}}\nonumber\\
&\leq 2d_\mr{sh}\exp\lrl{-\frac{||\hmo||_\mr{op}^{2}d_\mr{sh}^{1-2\eta}}{18\pi^3\Delta_\mc{O}^2}}.
}
Finally, using $\Delta_\mc{O}\leq 2 ||\mc{\hat{P}}_\mr{sh}\hmo\mc{\hat{P}}_\mr{sh}||_\mr{op}\leq 2 ||\hmo||_\mr{op}$, we obtain
\aln{\label{thm1}
\mbb{P}_U\lrl{\max_{|E_\alpha-E|, |E_\beta-E|\leq \Delta E}\lrv{\mc{O}_{\alpha\alpha}-\mc{O}_{\beta\beta}}> ||\hmo||_\mr{op} d_\mr{sh}^{-\eta}}\leq 2d_\mr{sh}\exp\lrl{-\frac{d_\mr{sh}^{1-2\eta}}{72\pi^3}},
}
which completes the proof of the inequality in (\ref{ineq}).\qed

The inequality in (\ref{ineq}) means that almost all (typical) $U$'s with respect to the Haar measure belong to $\mc{U}_{\{a_i\},\eta}$ in the thermodynamic limit if $0<\eta<\frac{1}{2}$ (see Fig. 1 in the main text).
Then, the typicality argument~\cite{Reimann15} asserts that even for a realistic pair of a Hamiltonian and an observable, we may expect that the corresponding $U$ satisfies $U\in\mc{U}_{\{a_i\},\eta}$.

\section{Proof of the \textbf{Proposition} in the main text}
We first show that a randomly chosen observable $\hat{G}=\sum_f G_f\hat{\Lambda}_f$ from $\mc{L}_N$ has eigenstates that are uniformly distributed with respect  to the Haar measure.
For an arbitrary $S^N\times S^N$ unitary transformation $\hat{R}$, we obtain
\aln{
\hat{R}\hat{G}\hat{R}^\dag&=\sum_{f=1}^{S^{2N}}G_f\hat{R}\hat{\Lambda}_f\hat{R}^\dag\nonumber\\ 
&=\sum_{f=1}^{S^{2N}}G_f\sum_{g=1}^{S^{2N}}\mc{R}_{fg}\hat{\Lambda}_g\nonumber\\ 
&=\sum_{f=1}^{S^{2N}}\tilde{G}_f\hat{\Lambda}_f,
}
where $\mc{R}_{fg}$ is defined through the operator expansion $\hat{R}\hat{\Lambda}_f\hat{R}^\dag=\sum_{g=1}^{S^{2N}}\mc{R}_{fg}\hat{\Lambda}_g$.
We define $\tilde{G}_f=\sum_{g=1}^{S^{2N}}G_g\mc{R}_{gf}$.
From the normalization condition of $\hat{\Lambda}_f$, namely $\Tr[\hat{\Lambda}_f\hat{\Lambda}_g]=S^N\delta_{fg}$, we obtain $\sum_h \mc{R}_{fh}\mc{R}_{gh}=\delta_{fg}$ by considering the operator expansion of $\hat{R}\hat{\Lambda}_f\hat{R}^\dag\hat{R}\hat{\Lambda}_g\hat{R}^\dag$.
Moreover, from the Hermiticity $\hat{\Lambda}_f^\dag=\hat{\Lambda}_f$, we obtain $\mc{R}_{fg}=\mc{R}_{fg}^*$ by considering the operator expansion of $(\hat{R}\hat{\Lambda}_f\hat{R}^\dag)^\dag$.
Thus, we can show that $\mc{R}$ is an $S^{2N}\times S^{2N}$ orthogonal matrix.
Then, if we pick up observables randomly from $\mc{L}_N$, the probabilities of choosing $\hat{G}$ and $\hat{R}\hat{G}\hat{R}^\dag$ are equal due to the invariance assumption $P(\vec{G})=P(\mc{R}\vec{G})$.
Consequently, if we diagonalize a randomly chosen $\hat{G}$, its eigenstates are uniformly distributed with respect to the unitary Haar measure.

Similarly to Eq. (\ref{ineq}), we can show the bound on the right-hand side in Eq. (7) in the main text.
By considering the unitary Haar measure for the entire Hilbert space, we obtain
\aln{
\mbb{P}_{\mc{L}_N}\lrl{\max_{\alpha,\beta}\lrv{\mc{O}_{\alpha\alpha}-\mc{O}_{\beta\beta}}> ||\hmo||_\mr{op} d_\mr{sh}^{-\eta}}\leq 2d\exp\lrl{-\frac{dd_\mr{sh}^{-2\eta}}{72\pi^3}}.
}
Since $\max_{\alpha,\beta}\lrv{\mc{O}_{\alpha\alpha}-\mc{O}_{\beta\beta}} \geq \max_{|E_\alpha-E|, |E_\beta-E|\leq \Delta E}\lrv{\mc{O}_{\alpha\alpha}-\mc{O}_{\beta\beta}}$, we obtain 
\aln{
\mbb{P}_{\mc{L}_N}\lrl{\max_{|E_\alpha-E|, |E_\beta-E|\leq \Delta E}\lrv{\mc{O}_{\alpha\alpha}-\mc{O}_{\beta\beta}}> ||\hmo||_\mr{op} d_\mr{sh}^{-\eta}}\leq 
\mbb{P}_{\mc{L}_N}\lrl{\max_{\alpha,\beta}\lrv{\mc{O}_{\alpha\alpha}-\mc{O}_{\beta\beta}}> ||\hmo||_\mr{op} d_\mr{sh}^{-\eta}}.
}
This completes the proof of Eq. (7) in the main text.\qed

\section{Eigenstate thermalization hypothesis for many-body correlations}
In this Appendix, we numerically show that the eigenstate thermalization hypothesis (ETH) is expected to hold true even for many-body correlations in a one-dimensional nonintegrable spin-1/2 system.

\begin{figure}
\begin{center}
\includegraphics[width=16cm]{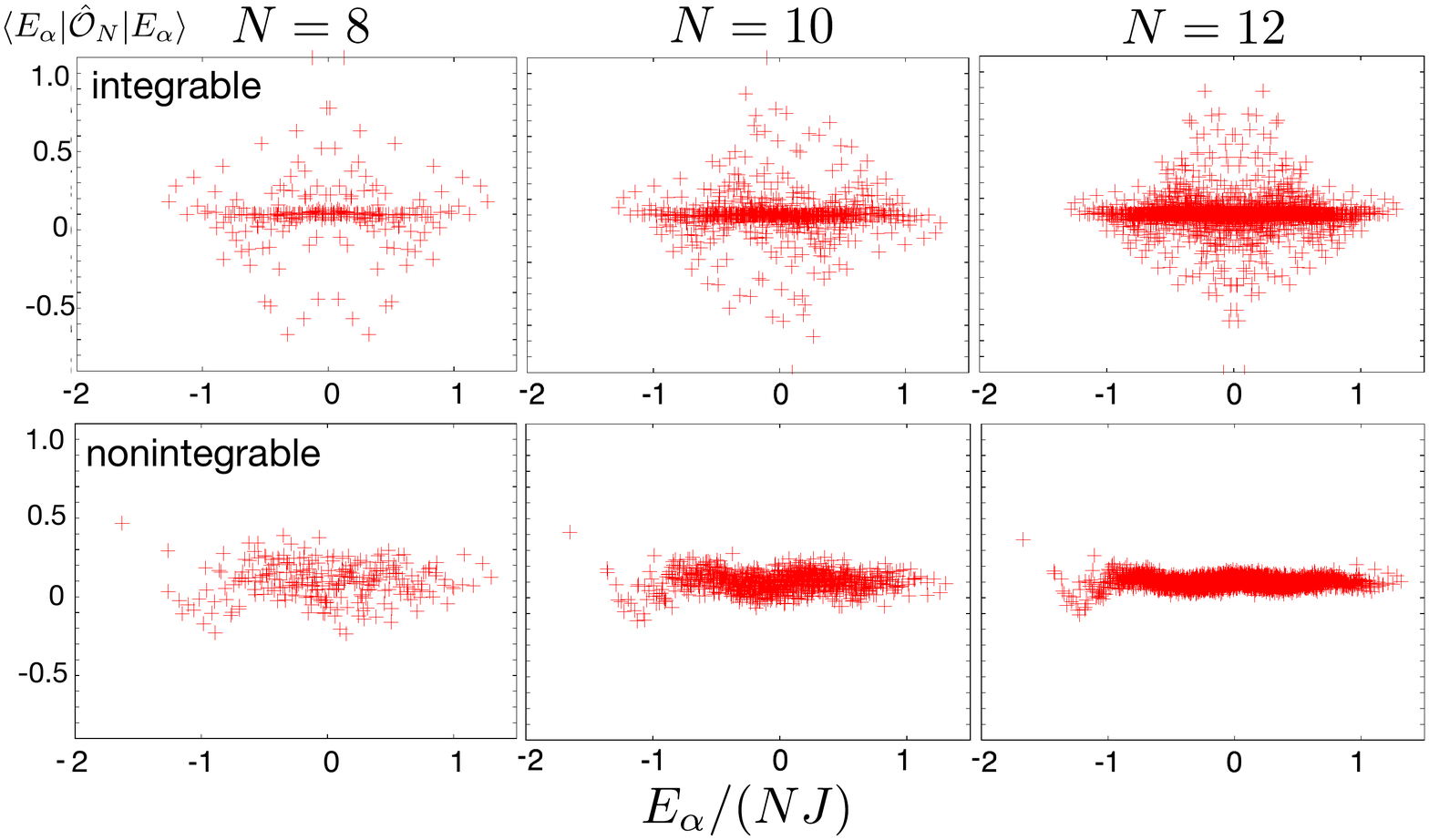}
\caption{
$N$-dependences of eigenstate expectation values (EEVs) of $\hmo_N$ for integrable (upper row) and nonintegrable (lower row) systems with $N=8$ (left), $N=10$ (middle), and $N=12$ (right).
The fluctuations of the EEVs decrease with increasing $N$ for nonintegrable systems, whereas
they remain large for integrable systems.
}
\label{maneth}
\end{center}
\end{figure}

In Fig. \ref{maneth}, we show the eigenstate expectation values (EEVs) $\braket{E_\alpha|\hat{\mc{O}}_N|E_\alpha}$  for integrable and nonintegrable systems.
Here, many-body correlations $\hat{\mc{O}}_N$ are defined as
\aln{\label{mbc}
\hat{\mc{O}}_N=\prod_{l=1}^N\hat{\sigma}_l^z,
}
where $\hat{\sigma}_l^z$ is the $z$ component of the Pauli operator at site $l$.

For an integrable system, we take a transverse-field Ising model with the open boundary condition whose Hamiltonian can be written as
\aln{
\hat{H}=-\sum_{l=1}^{N-1}J\hat{\sigma}_l^z\hat{\sigma}_{l+1}^z-\sum_{l=1}^Nh'\hat{\sigma}_l^x,
}
where we take $J=1$ and $h'=-1.05$.
For a nonintegrable system, we take a Hamiltonian 
\aln{
\hat{H}=-\sum_{l=1}^{N-1}J(1+\epsilon_l)\hat{\sigma}_l^z\hat{\sigma}_{l+1}^z-\sum_{l=1}^Nh'\hat{\sigma}_l^x-\sum_{l=1}^Nh\hat{\sigma}_l^z,
}
where $h=0.5$ and a random variable $\epsilon_l$ is uniformly chosen from $[-0.1,0.1]$ at each bond.

Figure \ref{maneth} shows that the fluctuations of the EEVs rapidly decrease with increasing $N$ for nonintegrable systems, whereas
they remain large for integrable systems.
This result implies that the ETH does (does not) hold true for nonintegrable (integrable) systems, even for many-body correlations given by Eq. (\ref{mbc}).
We note, however, that \textbf{Proposition} in the main text holds true regardless of whether the system is integrable or not.
From this proposition, we expect that the ETH holds true for more complex many-body observables even for integrable systems.

{
\section{Atypicality of most local observables}
}
We are often interested in the property of spatially local observables, rather than general few-body observables.
We can actually show atypicality of $\mc{O}_{\alpha\alpha}$ for most of such local observables in a manner similar to the case with \textbf{Theorem} in the main text, if the Hamiltonian is local and translationally invariant.
In the following, we consider spins on one-dimensional $N$ lattice sites ($x=1,\cdots, N$) for simplifying the notation (generalization to higher dimensions is straightforward).

We take a subsystem S$_l$ composed of $l$ neighboring lattices, which we call $x=1,2,\cdots, l$ (note that we can start from $x=1$ because we will consider a translation-invariant system below).
We assume that $l\:(\ll N)$ does not depend on $N$, which makes S$_l$ a local subsytem.
We define $\mc{H}_{\mr{S}_l}$ and $\mc{L}^\mr{loc}_{l}\equiv\mc{L}(\mc{H}_{\mr{S}_l})$ as the Hilbert space of S$_l$ and the operator space acting on S$_l$, respectively.
An orthonormal basis set for $\mc{L}^\mr{loc}_{l}$ can be written as $\mc{B}^\mr{loc}_l\equiv\lrm{\otimes_{x=1}^l \hat{\lambda}_x^{\mu_x}|0\leq \mu_x\leq S^2-1}\equiv\{\hat{\Lambda}^\mr{loc}_f\}_{f=1}^r$, where $r=S^{2l}$.
We choose the orthonormality condition as $\Tr_{\mr{S}_l}[\hat{\Lambda}^\mr{loc}_f\hat{\Lambda}^\mr{loc}_g]=S^{l}\delta_{fg}$.

Next we define random local observables from $\mc{L}^\mr{loc}_{l}$ as follows:
\definition[Randomly chosen local observables from $\mc{L}^\mr{loc}_l$]{
We take an observable $\hat{G}\in\mc{L}^\mr{loc}_{l}$ expressed as
\aln{
\hat{G}=\sum_{f=1}^rg_f\hat{\Lambda}^\mr{loc}_f,
}
where real variables $\vec{g}=(g_1,\cdots,g_f,\cdots,g_r)$ are randomly chosen according to an arbitrarily specified probability distribution $P^\mr{loc}(\vec{g})$.
When $P^\mr{loc}(\vec{g})$ is invariant under an arbitrary $r\times r$ orthogonal transformation, we call $\hat{G}$ an observable randomly chosen from $\mc{L}^\mr{loc}_l$.
}

We consider an $l'$-local, translation-invariant Hamiltonian $\hat{H}$ in the following sense. 
It is written as $\hat{H}=\sum_{i=1}^N\hat{h}_{i,l'}$, where $\hat{h}_{1,l'}$ only acts on $x=1,2,\cdots, l'$ and $\mc{T}[\hat{h}_{i,l'}]=\hat{h}_{i+1,l'}$ ($\mc{T}$ is the translation of one lattice).

Under this setup, we obtain the following theorem:
\theorem[Atypicality of most local observables]{
Let us consider an $l'$-local translation-invariant Hamiltonian, and assume that $N$ is sufficiently large and that $l \:(l'\leq l\ll N)$ is independent of $N$.
Suppose that we randomly choose an observable $\hmo=\sum_fg_f\hat{\Lambda}_f^\mr{loc}$ from $\mc{L}^\mr{loc}_l$, from which we obtain the corresponding ${\{a_i\}}$  and $U$.
Then,
\aln{\label{mthmloc}
\mbb{P}_{\mc{L}^\mr{loc}_l}[U\in \mc{U}_{\{a_i\},\eta}]\leq  \frac{\sqrt{\pi r}N||\hat{h}||_\mr{op}\Lambda }{2\Delta E}\frac{\Gamma\lrs{\frac{r}{2}}}{\Gamma\lrs{\frac{r-1}{2}}}d_\mr{sh}^{-\eta},
}
where $\mbb{P}_{\mc{L}^\mr{loc}_l}$ denotes a probability with respect to $P^\mr{loc}(\vec{g})$, and
 $\Lambda=\max_f||\hat{\Lambda}_f||_\mr{op}\leq S^{\frac{l}{2}}=r^{1/4}$.
When $||\hat{h}||_\mr{op}$ does not grow exponentially in $N$, the left-hand side vanishes for large $N$.
Note that the assumption of the scaling $\Delta E\propto N^{-p}$ is sufficient to bound the right-hand side.
}

\textit{Proof of \textbf{Theorem 1}}.
\:The proof is similar to that of \textbf{Theorem} in the main text.
We first note that $\hat{h}_{1,l'}\in\mc{L}^\mr{loc}_l$ and that $\hat{h}_{1,l'}$ satisfies the following condition:
\aln{\label{atyploc}
(\hat{h}_{1,l'})_{\gamma\gamma}-(\hat{h}_{1,l'})_{\delta\delta}&=\frac{1}{N}\sum_{i}[(\hat{h}_{i,l'})_{\gamma\gamma}-(\hat{h}_{i,l'})_{\delta\delta}]\nonumber\\
&=\frac{1}{N}[(\hat{H})_{\gamma\gamma}-(\hat{H})_{\delta\delta}]=\frac{2\Delta E}{N},
}
where $\gamma$ and $\delta$ are labels of maximum/minimum eigenvalues within the energy shell.
Define $\gamma_f=(\hat{\Lambda}^\mr{loc}_f)_{\gamma\gamma}-(\hat{\Lambda}^\mr{loc}_f)_{\delta\delta}$.
Then the expansion
$\hat{h}_{1,l'}=\sum_{f=1}^rh_f\hat{\Lambda}^\mr{loc}_f$
leads to
$\vec{h}\cdot\vec{\gamma}= \frac{2\Delta E}{N}$,
where $\vec{h}=(h_1,\cdots,h_{r})$ and $\vec{\gamma}=(\gamma_1,\cdots,\gamma_{r})$.
Since $|\vec{h}|=\sqrt{\frac{\Tr_{\mr{S}_l}[\hat{h}_{1,l'}^2]}{S^l}}\leq ||\hat{h}_{1,l'}||_\mr{op}\equiv ||\hat{h}||_\mr{op}$ (note that $||\hat{h}_{i,l'}||_\mr{op}=||\hat{h}||_\mr{op}$ for all $i$), we obtain
\aln{
|\vec{\gamma}|\geq \frac{2\Delta E}{N||\hat{h}||_\mr{op}}.
}

Next, we evaluate the left-hand side of Eq. (\ref{mthmloc}).
Since $\max_{|E_\alpha-E|, |E_\beta-E|\leq \Delta E}\lrv{\mc{O}_{\alpha\alpha}-\mc{O}_{\beta\beta}}\geq |\vec{g}\cdot\vec{\gamma}|$, we obtain
\aln{\label{atosukoshi}
\mbb{P}_{\mc{L}^\mr{loc}_l}\lrl{ \max_{|E_\alpha-E|, |E_\beta-E|\leq \Delta E}\lrv{\mc{O}_{\alpha\alpha}-\mc{O}_{\beta\beta}}\leq ||\hmo||_\mr{op}\epsilon}\leq \mbb{P}_{\mc{L}^\mr{loc}_l}\lrl{|\vec{g}\cdot\vec{\gamma}| \leq ||\hmo||_\mr{op}\epsilon}.
}

Next, denoting the angle between $\vec{g}$ and $\vec{\gamma}$ by $\phi$, we obtain
\aln{\label{owari}
 \mbb{P}_{\mc{L}^\mr{loc}_l}\lrl{|\vec{g}\cdot\vec{\gamma}| \leq ||\hmo||_\mr{op}\epsilon}
&\leq \mbb{P}_{\mc{L}^\mr{loc}_l}\lrl{|\cos\phi| \leq \frac{\sqrt{r}N||\hat{h}||_\mr{op}\Lambda \epsilon}{2\Delta E} }\nonumber\\
&\leq \frac{\sqrt{\pi r}N||\hat{h}||_\mr{op}\Lambda \epsilon}{2\Delta E}\frac{\Gamma\lrs{\frac{r}{2}}}{\Gamma\lrs{\frac{r-1}{2}}}.
}
Here, in deriving the second line, we have used
$
||\hmo||_\mr{op}\leq\Lambda |\vec{g}|\sqrt{r}
$
and the invariance condition of $P^\mr{loc}(\vec{g})$ (see also the main text).
Combining (\ref{atosukoshi}) and (\ref{owari}), and taking $\epsilon=d_\mr{sh}^{-\eta}$, the left-hand side of (\ref{atosukoshi}) becomes $\mbb{P}_{\mc{L}^\mr{loc}_l}[U\in \mc{U}_{\{a_i\},\eta}]$, which completes the proof of the theorem.\qed

From (\ref{atosukoshi}) and (\ref{owari}), we can further estimate how slowly the maximum variation of diagonal matrix elements decays.
To do this, we take $\Delta E=||\hat{h}||_\mr{op}N^{-p}\: (-1<p)$.
Since $\frac{\sqrt{\pi r}\Lambda}{2}\frac{\Gamma\lrs{\frac{r}{2}}}{\Gamma\lrs{\frac{r-1}{2}}}$
does not depend on $N$, we can take $\epsilon =N^{-z}\:(z> 1+p)$ for the left-hand side of (\ref{atosukoshi}) to vanish in the thermodynamic limit.
We note that this does not contradict the result of the weak ETH~\cite{Biroli10}.
The weak ETH states that the variance of $\mc{O}_{\alpha\alpha}$ within the energy shell decreases faster than $N^{-1}$ if the equilibrium correlation function of $\hat{\mc{O}}$ rapidly decreases with the distance.
Since the maximum variation is always larger than the variance, our result and the weak ETH can be compatible (indeed, the weak ETH is known to hold for integrable systems, for which the strong ETH breaks down).

\section{Off-diagonal matrix elements}
We here analyze off-diagonal matrix elements.
For energy eigenstates 
$\ket{E_\alpha},\ket{E_\beta}\in\mc{H}_\mr{sh}$,
we obtain
$E-\Delta E \leq \frac{E_\alpha+E_\beta}{2}\leq E+\Delta E$ and $-2\Delta E\leq E_\beta-E_\alpha =\omega_{\alpha\beta}\leq 2\Delta E$, where we set $\hbar=1$.
If the off-diagonal matrix elements have almost the same order of magnitude over the energy shell $\mathcal{H}_\mathrm{sh}$, no characteristic timescale of thermalization for $\hmo$ should appear after $\sim\frac{1}{2\Delta E}$.

\begin{figure}
\begin{center}
\includegraphics[width=15cm]{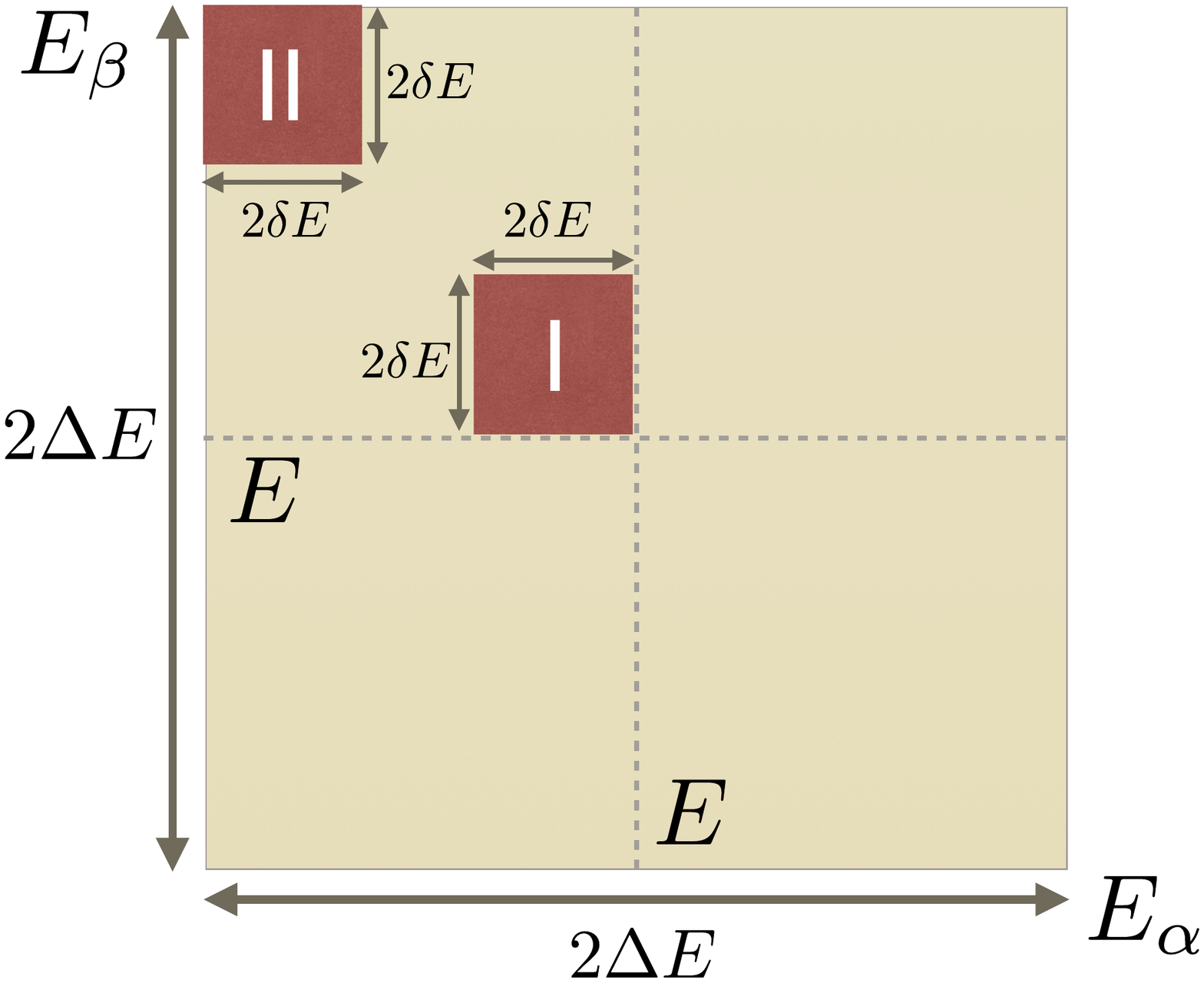}
\caption{
Region I and II for off-diagonal matrix elements.
Region I satisfies $E-2\delta E<E_\alpha<E$ and $E\leq E_\beta\leq E+2\delta E$, where $\delta E (\ll \Delta E)$ is a small constant.
Region II satisfies $E-\Delta E<E_\alpha<E-\Delta E+2\delta E$ and $E+\Delta E-2\delta E\leq E_\beta\leq E+\Delta E$.
}
\label{offd}
\end{center}
\end{figure}

We define the following regions I and II for matrix elements (see Fig.\ref{offd}).
Region I satisfies $E-2\delta E<E_\alpha<E, E\leq E_\beta\leq E+2\delta E$, where $\delta E (\ll \Delta E)$ is a small constant.
We denote the numbers of energy eigenstates $\ket{E_\alpha}$ and $\ket{E_\beta}$ that satisfy this condition by $p_\mr{I}$ and $q_\mr{I}$, respectively.
Then the total number of relevant matrix elements is $p_\mr{I}q_\mr{I}$.
Note that
$0< \omega_{\alpha\beta}<4\delta E$ is satisfied.
Similarly, region II satisfies $E-\Delta E<E_\alpha<E-\Delta E+2\delta E$ and $E+\Delta E-2\delta E\leq E_\beta\leq E+\Delta E$.
We denote the numbers of energy eigenstates $\ket{E_\alpha}$ and $\ket{E_\beta}$ that satisfy this condition by $p_\mr{II}$ and $q_\mr{II}$, respectively.
Then the total number of relevant matrix elements is $p_\mr{II}q_\mr{II}$.
Note that the condition 
$2\Delta E-4\delta E< \omega_{\alpha\beta}<2\Delta E$ is satisfied.
 
As a quantity of our interest, we define the spectral average of off-diagonal matrix elements for each region as follows:
\aln{
\braket{\mc{O}_\mr{off}^2}_\mr{I}&=\frac{1}{p_\mr{I}q_\mr{I}}\sum_\mr{I} |\mc{O}_{\alpha\beta}|^2,\\
\braket{\mc{O}_\mr{off}^2}_\mr{II}&=\frac{1}{p_\mr{II}q_\mr{II}}\sum_\mr{II} |\mc{O}_{\alpha\beta}|^2.
}
Here each sum is taken over all matrix elements in I or II.
In the following discussions, we consider $d_\mr{sh}\braket{\mc{O}_\mr{off}^2}_\mr{I/II}$, since $\braket{\mc{O}_\mr{off}^2}_\mr{I/II}$ is expected to decrease as $\sim d_\mr{sh}^{-1}$~\cite{Beugeling15} (see Ref.~\cite{Brody81} for an exception).

First, we can prove a statement similar to Eq. (\ref{ineq}) for the off-diagonal spectral average.
 For this purpose, {we recall that matrix elements $\mc{O}_{\alpha\beta}$ within the energy shell are determined by a unitary matrix $U$ through $\mc{O}_{\alpha\beta}=\sum_{i=1}^{d_\mr{sh}}a_iU_{\alpha i}U_{\beta i}^*$ for fixed eigenvalues $\{a_i\}$. Similarly, $\braket{\mc{O}_\mr{off}^2}_\mr{I}$ and $\braket{\mc{O}_\mr{off}^2}_\mr{II}$ are also calculated from $U$.}
 Then we have the following proposition.

\proposition[Typicality with respect to the Haar measure]{
{We define} $\mc{U}'_{\{a_i\},\eta}$ {as} a set of {all} $U$'s that lead to the inequality $\lrv{d_\mr{sh}\braket{\mc{O}_\mr{off}^2}_\mr{I}-d_\mr{sh}\braket{\mc{O}_\mr{off}^2}_\mr{II}}\leq||\hmo||_\mr{op}^2d_\mr{sh}^{-\eta}$ for given ${\{a_i\}}$, where $\eta>0$.
We then have 
\aln{\label{maina}
\mbb{P}_U&\lrl{U\notin\mc{U}'_{\{a_i\},\eta}
 }\leq\frac{4B_\mr{I}}{d_\mr{sh}^{-2\eta} d_\mr{I}}+\frac{4B_\mr{II}}{d_\mr{sh}^{-2\eta} d_\mr{II}},
}
 where $B_\mr{I}$ and $B_\mr{II}$ are some constants, $d_\mr{I}=\min\{p_\mr{I},q_\mr{I}\}$, and $d_\mr{II}=\min\{p_\mr{II},q_\mr{II}\}$.
If $d_\mr{I},d_\mr{II}\gg d_\mr{sh}^{2\eta}$ in the thermodynamic limit, the right-hand side vanishes in this limit.
This means that the variations of typical magnitudes of off-diagonal matrix elements decrease exponentially within an energy shell.

\textit{Proof.}
First we show
\aln{\label{ine}
\mbb{P}_U\lrl{\lrv{d_\mr{sh}\braket{\mc{O}_\mr{off}^2}_\mr{I}-d_\mr{sh}\av{|\mc{O}_{\alpha\beta}|^2}}>||\hmo||_\mr{op}^2d_\mr{sh}^{-\eta}}\leq\frac{B_\mr{I}}{d_\mr{sh}^{-2\eta} d_\mr{I}}
}
and a similar inequality for II, where the overline denotes the average with respect to the Haar measure.
We begin by calculating the ensemble average of the spectral average over the Haar measure:
\aln{
\av{d_\mr{sh}\braket{\mc{O}_\mr{off}^2}_\mr{I}}&=\frac{d_\mr{sh}}{p_\mr{I}q_\mr{I}}\sum_\mr{I} \av{|\mc{O}_{\alpha\beta}|^2}\nonumber\\
&= d_\mr{sh}\av{|\mc{O}_{\alpha\beta}|^2}.
}
Here we have used the fact that $\av{|\mc{O}_{\alpha\beta}|^2}$ is independent of $\alpha$ and $\beta$ in the energy shell.
Next we consider the ensemble variance of the spectral average.
When $d_\mr{I}(<d_\mr{sh})$ is large enough, we obtain
\aln{\label{bubun}
\av{\lrs{d_\mr{sh}\braket{\mc{O}_\mr{off}^2}_\mr{I}-\av{d_\mr{sh}\braket{\mc{O}_\mr{off}^2}_\mr{I}}}^2}\leq B_\mr{I}||\hmo||^4_\mr{op}d_\mr{I}^{-1},
}
where $B_I$ is a constant.
The proof is given in a similar manner as in Ref.~\cite{Brody81}.
By Chebyshev's inequality, we obtain Eq. (\ref{ine}).
A similar proof can be made for the region II.

Now we show \textbf{Proposition 1}. Since
\aln{
\lrv{d_\mr{sh}\braket{\mc{O}_\mr{off}^2}_\mr{I}-d_\mr{sh}\braket{\mc{O}_\mr{off}^2}_\mr{II}}\leq 2\max\lrm{\lrv{d_\mr{sh}\braket{\mc{O}_\mr{off}^2}_\mr{I}-d_\mr{sh}\av{|\mc{O}_{\alpha\beta}|^2} },\lrv{d_\mr{sh}\braket{\mc{O}_\mr{off}^2}_\mr{II}-d_\mr{sh}\av{|\mc{O}_{\alpha\beta}|^2} }},
}
 we obtain
\aln{\label{shomei}
\mbb{P}_U&\lrl{\lrv{d_\mr{sh}\braket{\mc{O}_\mr{off}^2}_\mr{I}-d_\mr{sh}\braket{\mc{O}_\mr{off}^2}_\mr{II}}>||\hmo||_\mr{op}^2d_\mr{sh}^{-\eta}
 }\leq\nonumber\\
&\mbb{P}_U\lrl{ 2\max\lrm{\lrv{d_\mr{sh}\braket{\mc{O}_\mr{off}^2}_\mr{I}-d_\mr{sh}\av{|\mc{O}_{\alpha\beta}|^2} },\lrv{d_\mr{sh}\braket{\mc{O}_\mr{off}^2}_\mr{II}-d_\mr{sh}\av{|\mc{O}_{\alpha\beta}|^2} }
}>||\hmo||_\mr{op}^2d_\mr{sh}^{-\eta}}\nonumber\\
&\leq\mbb{P}_U\lrl{\lrv{d_\mr{sh}\braket{\mc{O}_\mr{off}^2}_\mr{I}-d_\mr{sh}\av{|\mc{O}_{\alpha\beta}|^2} }>||\hmo||_\mr{op}^2d_\mr{sh}^{-\eta}/2}+
\mbb{P}_U\lrl{\lrv{d_\mr{sh}\braket{\mc{O}_\mr{off}^2}_\mr{II}-d_\mr{sh}\av{|\mc{O}_{\alpha\beta}|^2} }>||\hmo||_\mr{op}^2d_\mr{sh}^{-\eta}/2}\nonumber\\
&\leq\frac{4B_\mr{I}}{d_\mr{sh}^{-2\eta} d_\mr{I}}+\frac{4B_\mr{II}}{d_\mr{sh}^{-2\eta} d_\mr{II}},
}
where we have used $\mbb{P}[\max\{a, b\}>\epsilon]\leq\mbb{P}[a>\epsilon]+\mbb{P}[b>\epsilon]$.\qed

Just as we have done for diagonal matrix elements, we can show that most few-body observables are atypical in the sense that the corresponding $U$ does not belong to $\mc{U}'_{\{a_i\},\eta}$. This can be stated in the form of the following theorem.
\theorem{
Suppose that there exists an observable $\hat{A}\in \mc{L}_m$ such that $d_\mr{sh}\braket{{A}_\mr{off}^2}_\mr{II}$ and $||\hat{A}||_\mr{op}^{-1}$ decrease no faster than polynomials in $N$.
We also assume that the Hamiltonian is a $k$-body operator ($k\leq m$) and $||\hat{H}||_\mr{op}$ does not increase exponentially in $N$.

For sufficiently small $\delta E$, we can show that
\aln{
\mbb{P}_{\mc{L}_m}\lrl{U\in \mc{U}'_{\{a_i\},\eta}
}\leq  \sqrt{\frac{2\pi }{\xi_\mr{od}}}\frac{n^\frac{3}{4}||\hat{B}||_\mr{op}\Lambda\Gamma\lrs{\frac{n}{2}}}{\Gamma\lrs{\frac{n-1}{2}}}
d_\mr{sh}^{-\frac{\eta}{2}}.
}
Here
$\hat{B}$ is either $\hat{A}$ or $i[\hat{H},\hat{A}]$, 
and $\xi_\mr{od}$ is some constant that decreases no faster than polynomials in $N$.
The right-hand side vanishes in the thermodynamic limit when $m$ does not depend on $N$.
}
This means the following: the typical homogeneous variance inside the energy shell, which is predicted by the unitary Haar measure, cannot be observed for the off-diagonal matrix elements of most few-body observables.

\textit{Proof.}
We first seek for an observable $\hat{B}\in\mc{L}_m$ that satisfies the following condition:
\aln{\label{atyp2}
\lrv{d_\mr{sh}\braket{B_\mr{off}^2}_\mr{I}-d_\mr{sh}\braket{B_\mr{off}^2}_\mr{II}}\geq \xi_\mr{od},
}
where $\xi_\mr{od}$ does not decrease faster than polynomials in $N$.
We show that either $\hat{A}$ or $i[\hat{H},\hat{A}]$ satisfy this condition.

Let us take a positive constant $c$.
If $|d_\mr{sh}\braket{{A}_\mr{off}^2}_\mr{II}-d_\mr{sh}\braket{{A}_\mr{off}^2}_\mr{I}|\geq c$, we can take $\hat{B}=\hat{A}$ and $\xi_\mr{od}=c$; otherwise, we take $\hat{B}=i[\hat{H},\hat{A}]$.
Then
\aln{
\lrv{d_\mr{sh}\braket{{B}_\mr{off}^2}_\mr{I}-d_\mr{sh}\braket{{B}_\mr{off}^2}_\mr{II}}
&=\lrv{\frac{d_\mr{sh}}{p_\mr{I}q_\mr{I}}\sum_\mr{I} (E_\alpha-E_\beta)^2|{A}_{\alpha\beta}|^2-\frac{d_\mr{sh}}{p_\mr{II}q_\mr{II}}\sum_\mr{II}(E_\alpha-E_\beta)^2 |{A}_{\alpha\beta}|^2}.
}
We note that 
\aln{
\frac{d_\mr{sh}}{p_\mr{II}q_\mr{II}}\sum_\mr{II}(E_\alpha-E_\beta)^2 |{A}_{\alpha\beta}|^2&\geq 4(\Delta E-2\delta E)^2 d_\mr{sh}\braket{{A}_\mr{off}^2}_\mr{II},\nonumber\\
\frac{d}{p_\mr{I}q_\mr{I}}\sum_\mr{I} (E_\alpha-E_\beta)^2|{A}_{\alpha\beta}|^2&\leq 16\delta E^2 d_\mr{sh}\braket{{A}_\mr{off}^2}_\mr{I}
\leq 16\delta E^2(c+ d_\mr{sh}\braket{{A}_\mr{off}^2}_\mr{II}).
}
Thus, if we take $\delta E$ such that
\aln{
\delta E^2\leq\frac{\Delta E^2d_\mr{sh}\braket{{A}_\mr{off}^2}_\mr{II}}{64(c+ d_\mr{sh}\braket{{A}_\mr{off}^2}_\mr{II})}\lrs{\leq \frac{\Delta E^2}{64}}
}
for a fixed $\Delta E$,
we obtain
\aln{
\lrv{d_\mr{sh}\braket{{B}_\mr{off}^2}_\mr{I}-d\braket{{B}_\mr{off}^2}_\mr{II}}
&=\lrv{\frac{d_\mr{sh}}{p_\mr{I}q_\mr{I}}\sum_\mr{I} (E_\alpha-E_\beta)^2|{A}_{\alpha\beta}|^2-\frac{d_\mr{sh}}{p_\mr{II}q_\mr{II}}\sum_\mr{II}(E_\alpha-E_\beta)^2 |{A}_{\alpha\beta}|^2}\nonumber\\
&\geq 4(\Delta E-2\delta E)^2 d_\mr{sh}\braket{{A}_\mr{off}^2}_\mr{II}-16\delta E^2 (c+ d_\mr{sh}\braket{{A}_\mr{off}^2}_\mr{II})\nonumber\\
&\geq \frac{9}{4}{\Delta E^2} d_\mr{sh}\braket{{A}_\mr{off}^2}_\mr{II}-\frac{1}{4}{\Delta E^2} d_\mr{sh}\braket{{A}_\mr{off}^2}_\mr{II}\nonumber\\
&= 2{\Delta E^2} d_\mr{sh}\braket{{A}_\mr{off}^2}_\mr{II}.
}
Since the last term decreases no faster than polynomials in $N$, we can take $\xi_\mr{od}= 2{\Delta E^2} d_\mr{sh}\braket{{A}_\mr{off}^2}_\mr{II}$.

We note that for $\hat{B}=i[\hat{H},\hat{A}]$,  $\sqrt{d_\mr{sh}\braket{B_\mr{off}^2}_\mr{II}}-\sqrt{d_\mr{sh}\braket{B_\mr{off}^2}_\mr{I}}\propto \Delta E$ approximately holds for sufficiently small $\delta E$.
Thus, for such a few-body operator, we do not see the plateau-like structure of off-diagonal matrix elements suggested in nonintegrable systems~\cite{DAlessio16} even for small $\Delta E$.

Now we give the proof of the atypicality of most few-body observables.
Let us define 
\aln{
2\mc{Z}_{fg}&=d_\mr{sh}\braket{\hat{\Lambda}_f:\hat{\Lambda}_g}_\mr{II}-d_\mr{sh}\braket{\hat{\Lambda}_f:\hat{\Lambda}_g}_\mr{I}+\mr{c.c.},\\
\braket{\hat{\Lambda}_f:\hat{\Lambda}_g}_\mr{II}&=\frac{1}{p_\mr{II}q_\mr{II}}\sum_\mr{II}(\hat{\Lambda}_f)_{\alpha\beta}(\hat{\Lambda}_g)_{\beta\alpha},\\
\braket{\hat{\Lambda}_f:\hat{\Lambda}_g}_\mr{I}&=\frac{1}{p_\mr{I}q_\mr{I}}\sum_\mr{I}(\hat{\Lambda}_f)_{\alpha\beta}(\hat{\Lambda}_g)_{\beta\alpha}.
}
Then
$
\hat{B}=\sum_fB_f\hat{\Lambda}_f
$
leads to
\aln{
\lrv{\sum_{fg}B_fB_g\mc{Z}_{fg}}\geq \xi_\mr{od}.
}
Here, $\mc{Z}_{fg}$ is real and symmetric, and can be diagonalized by an orthogonal transformation $\mc{R}$.
By writing $\mc{R}\mc{Z}\mc{R}^T=\mr{diag}(D_1,\cdots, D_f,\cdots ,D_{n})$ with $D_1\leq \cdots\leq  D_n$ and $\vec{B'}=\mc{R}\vec{B}$, we have
\aln{
\lrv{\sum_fB_f'^2D_f}\geq \xi_\mr{od}.
}
We note that 
\aln{
\sum_fB_f'^2\sqrt{\sum_fD_f^2}\geq \sqrt{\sum_fB_f'^4\sum_fD_f^2}\geq \lrv{\sum_fB_f'^2D_f}\geq \xi_\mr{od},
}
which is obtained by using $\lrs{\sum_fB_f'^2}^2\geq\sum_fB_f'^4$ and the Cauchy-Schwarz inequality.
We also note that
\aln{
\sum_fB_f'^2=\sum_fB_f^2=\frac{\Tr[\hat{B}^2]}{S^{N}}\leq ||\hat{B}||_\mr{op}^2.
}
Then
\aln{
|\vec{D}|&=\sqrt{\sum_fD_f^2}\geq \frac{\xi_\mr{od}}{||\hat{B}||_\mr{op}^2},
}
 where $\vec{D}=(D_1,\cdots,D_n)$.
 We also define  $D_M:=\max_f|D_f|\geq \frac{\xi_\mr{od}}{\sqrt{n}||\hat{B}||_\mr{op}^2}$.
 
We now show that
 \aln{
\mbb{P}_{\mc{L}_m}\lrl{\lrv{d\braket{\mc{O}_\mr{off}^2}_I-d\braket{\mc{O}_\mr{off}^2}_{II}}\leq ||\hmo||_\mr{op}^2d_\mr{sh}^{-\eta}}=
\mbb{P}_{\mc{L}_m}\lrl{\lrv{\sum_{fg}G_fG_g\mc{Z}_{fg}}\leq ||\hmo||_\mr{op}^2d_\mr{sh}^{-\eta}}
}
is small {(note that $\hmo=\sum_fG_f\hat{\Lambda}_f$)}. 
Since the probability is invariant under the orthogonal transformation $\mc{R}$, we have
 \aln{\label{hebo}
\mbb{P}_{\mc{L}_m}\lrl{\lrv{\sum_{fg}G_fG_g\mc{Z}_{fg}}\leq ||\hmo||_\mr{op}^2d_\mr{sh}^{-\eta}}=
\mbb{P}_{\mc{L}_m}\lrl{\lrv{\sum_{f}G_f^2D_{f}}\leq ||\hmo||_\mr{op}^2d_\mr{sh}^{-\eta}}.
}
To evaluate the right-hand side, we first fix $|\vec{G}|$ and consider the probability on the hypersphere $\sum_f G_f^2=|\vec{G}|^2$.
We note that
\aln{
||\hmo||_\mr{op}^2&\leq \lrs{\sum_{f=1}^n|G_f|\cdot||\hat{\Lambda}_f||_\mr{op}}^2\nonumber\\
&\leq \Lambda^2\lrs{\sum_{f=1}^n|\vec{G}_f|}^2\nonumber\\
&\leq \Lambda^2|\vec{G}|^2n,
}
where the last inequality comes from the Cauchy-Schwarz inequality.
Thus, the right-hand side in Eq. (\ref{hebo}) is further bounded from above by
\aln{\label{mtm}
\mbb{P}_{\mc{L}_m}\lrl{-|\vec{G}|^2\delta\leq \sum_{f=1}^nG_f^2D_{f}\leq |\vec{G}|^2\delta},
}
 where $\delta =\Lambda^2nd_\mr{sh}^{-\eta}$ decreases exponentially as a function of $N$.
 
Without loss of generality, we can assume $D_1=\min_fD_f=-D_M$ (the following discussion holds true for the case with $D_n=\max_fD_f=D_M$).
Then, Eq. (\ref{mtm}) is equivalent to
\aln{\label{daen}
\mbb{P}_{\mc{L}_m}\lrl{|\vec{G}|^2(-\delta+D_M)\leq \sum_{f=2}^nG_f^2E_{f}\leq |\vec{G}|^2(\delta+D_M)}
}
under the constraint 
\aln{\label{en}
\sum_{f=1}^n G_f^2=|\vec{G}|^2,
}
where $E_f=D_f+D_M>0$ for $f\geq 2$.

\begin{figure}
\begin{center}
\includegraphics[width=15cm]{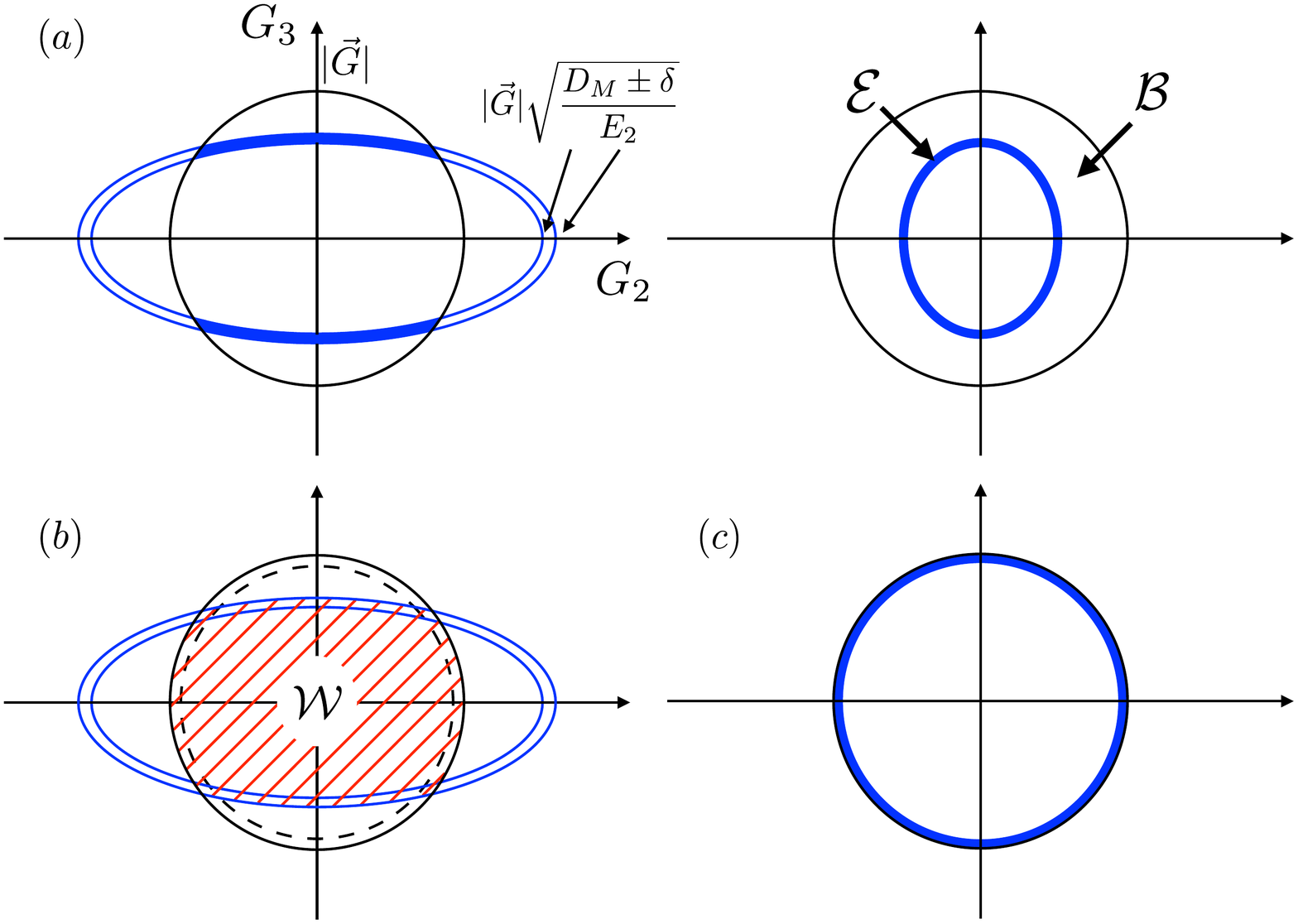}
\caption{
(a) Two possible configurations of overlaps of an $(n-1)$-dimensional ball ($\mc{B}$) and an $(n-1)$-dimensional thin elliptic hypercylinder shell ($\mc{E}$) for $n=3$.
(b) Evaluation of the volume of the overlap $\mc{N}'$ via $\mc{W}$ (Eq. (\ref{nazoieq})). 
(c) Circular configuration for which both $\mc{W}$ and $\mc{N}'$ are maximized.
In this case, $\mc{N}$ is also maximized.
}
\label{overlap}
\end{center}
\end{figure}

Equations (\ref{daen}) and (\ref{en}) allow a geometrical interpretation that we should evaluate an overlap of the $(n-1)$-dimensional hypersphere and the $n$-dimensional thin  elliptic hypercylinder shell.
For fixed $|\vec{G}|$, the volume of the overlap can be evaluated by integrating out $G_1$:
\aln{
\mc{N}=\int_{\mc{B} \cap \mc{E}}\frac{dG_2\cdots dG_n}{\sqrt{|\vec{G}|^2-G_2^2-\cdots-G_n^2}},
}
where $\mc{B}$ denotes an $(n-1)$-dimensional ball with the radius $|\vec{G}|$ and $\mc{E}$ denotes an $(n-1)$-dimensional thin elliptic shell (see Figure \ref{overlap}).

We consider those configurations of $\vec{E}$ which maximize the overlap volume $\mc{N}$.
We first note that the quantity
\aln{
\mc{N}'=\int_{\mc{B} \cap \mc{E}}dG_2\cdots dG_n,
}
satisfies
\aln{\label{nazoieq}
\mc{N}'\leq \lrl{1-\lrs{\frac{D_M-\delta}{D_M+\delta}}^{\frac{n-1}{2}}}\mc{W},
}
where $\mc{W}$ is the overlap volume between $\mc{B}$ and the region $\lrm{\sum_{f=2}^nG_f^2E_f\leq |\vec{G}|^2(\delta+D_M)}$ (see Fig. \ref{overlap}(b)).
We note that $\mc{W}$ is maximized when $\mc{B}\subset\lrm{\sum_{f=2}^nG_f^2E_f\leq |\vec{G}|^2(\delta+D_M)}$.
Furthermore, the equality of Eq. (\ref{nazoieq}) is attained when $\lrm{\sum_{f=2}^nG_f^2E_f\leq |\vec{G}|^2(\delta+D_M)}\subset\mc{B}$.
Thus, $\mc{N}'$ is maximized when $\mc{B}=\lrm{\sum_{f=2}^nG_f^2E_f\leq |\vec{G}|^2(\delta+D_M)}$ (see Fig. \ref{overlap}(c)).
In this case, $\mc{N}$ is also maximized, since the factor $\sqrt{|\vec{G}|^2-G_2^2-\cdots-G_n^2}$ becomes smaller as $(G_2,\cdots,G_n)$ approaches the edge of $\mc{B}$. 
Thus, Eq. (\ref{daen}) is maximized when $\vec{E}=(D_M+\delta, \cdots,D_M+\delta)$.

Going back to the original problem, we obtain
\aln{
\mbb{P}_{\mc{L}_m}\lrl{-|\vec{G}|^2\delta\leq \sum_{f=1}^nG_f^2D_{f}\leq |\vec{G}|^2\delta}
\leq \mbb{P}_{\mc{L}_m}\lrl{-|\vec{G}|^2\delta\leq \sum_{f=1}^nG_f^2D^\mr{M}_{f}\leq |\vec{G}|^2\delta}
}
for any fixed $\vec{D}$, where $\vec{D}^\mr{M}=(-D_M, \delta,\cdots,\delta)$.
If we denote the angle between the $\vec{G}$-axis and the $G_1$-axis by $\theta$, we obtain
\aln{
\mbb{P}_{\mc{L}_m}\lrl{-|\vec{G}|^2\delta\leq \sum_{f=1}^nG_f^2D^\mr{M}_{f}\leq |\vec{G}|^2\delta}&=
\mbb{P}_{\mc{L}_m}\lrl{-\delta\leq -D_M\cos^2\theta+\delta \sin^2\theta\leq \delta}\\
&\leq \mbb{P}_{\mc{L}_m}\lrl{-\sqrt{\frac{2\delta}{D_M}} \leq \cos\theta\leq \sqrt{\frac{2\delta}{D_M}}}\\
&\leq \sqrt{\frac{2\pi }{\xi_\mr{od}}}n^{\frac{3}{4}}||\hat{B}||_\mr{op}\Lambda d_\mr{sh}^{-\frac{\eta}{2}}\frac{\Gamma\lrs{\frac{n}{2}}}{\Gamma\lrs{\frac{n-1}{2}}},
}
which completes the proof of \textbf{Theorem 2.}\qed

Finally, we consider observables randomly chosen from $\mc{L}_{m=N}$, which include many-body observables.
Then, most of them satisfy $U\in \mc{U}'_{\{a_i\},\eta}$ as stated in the following proposition.
\proposition[]{
Let $\mbb{P}_{\mc{L}_m}$ be a probability with respect to $P(\vec{G})$.
If we take $m=N$, the following inequality holds.
\aln{\label{zenbu}
\mbb{P}_{\mc{L}_N}&\lrl{U\notin\mc{U}'_{\{a_i\},\eta}
 }\leq\frac{4B_\mr{I}d_\mr{sh}^{2+2\eta}}{d^2 d_\mr{I}}+\frac{4B_\mr{II}d_\mr{sh}^{2+2\eta}}{d^2 d_\mr{II}}.
}
The right-hand side vanishes when $\frac{d_\mr{sh}^{2+2\eta}}{d^2d_\mr{I}},\frac{d_\mr{sh}^{2+2\eta}}{d^2d_\mr{II}}\ll 1$ for sufficiently large $N$. 
}
The proof goes similarly to that of \textbf{Proposition} in the main text.
We can show that $U$ distributes uniformly over the unitary Haar measure over the entire Hilbert space, and the same method for proving \textbf{Proposition 1} can be used.
That is, $d_\mr{sh}$ in the left-hand side of Eq. (\ref{bubun}) can be replaced by $d$.
Consequently, we can use the Chebyshev's inequality to
$\mbb{P}_{\mc{L}_N}\lrl{\lrv{d\braket{\mc{O}_\mr{off}^2}_\mr{I}-d\braket{\mc{O}_\mr{off}^2}_\mr{II}}>||\hmo||_\mr{op}^2d d_\mr{sh}^{-1-\eta} }$, which leads to Eq. (\ref{zenbu}).

\bibliography{../../../../refer_them}